\documentclass{article}
\usepackage[utf8]{inputenc}
\usepackage{graphicx, amssymb, pifont, amsmath}
\usepackage{hyperref}
\usepackage{caption}
\usepackage{subcaption}
\usepackage[affil-it]{authblk}
\usepackage{geometry}

\usepackage{color}

\title{Spaces of innovation and venture formation: the case of biotech in the United Kingdom}
\author[1,2]{Francesco Marzolla}
\author[3]{Przemysław Nowak}
\author[4]{Rohit Sahasrabuddhe}
\author[5]{Chakresh Singh}
\author[1,2,6]{Matteo Straccamore}
\author[7]{Erik Zhivkoplias}
\author[8]{Elsa Arcaute}

\affil[1]{SONY Computer Science Laboratories, Rome, Via Panisperna 89A, 00184 Rome, Italy}
\affil[2]{Sapienza Univ. of Rome, Physics Dept., Piazzale Aldo Moro 2, 00185, Rome, Italy}
\affil[3]{Warsaw University of Technology, Faculty of Physics, ul.~Koszykowa 75, 00-662 Warsaw, Poland}
\affil[4]{Mathematical Institute, University of Oxford, United Kingdom}

\affil[5]{Universit$\acute{e}$ Paris Cit$\acute{e}$, Inserm, System Engineering and
Evolution Dynamics, F-75004, Paris, France.}
\affil[6]{Centro Ricerche Enrico Fermi, Via Panisperna 89/A, 00184, Rome, Italy}
\affil[7]{Stockholm Resilience Centre, Stockholm University, 106 91, Stockholm, Sweden}
\affil[8]{Centre for Advanced Spatial Analysis, University College London, United Kingdom}

\begin{document}

\maketitle

\begin{abstract}
Patents serve as valuable indicators of innovation and provide insights into the spaces of innovation and venture formation within geographic regions. In this study, we utilise patent data to examine the dynamics of innovation and venture formation in the biotech sector across the United Kingdom (UK). By analysing patents, we identify key regions that drive biotech innovation in the UK. Our findings highlight the crucial role of biotech incubators in facilitating knowledge exchange between scientific research and industry. However, we observe that the incubators themselves do not significantly contribute to the diversity of innovations which might be due to the underlying effect of geographic proximity on the influences and impact of the patents. These insights contribute to our understanding of the historical development and future prospects of the biotech sector in the UK, emphasising the importance of promoting innovation diversity and fostering inclusive enterprise for achieving equitable economic growth.

\end{abstract}
\section*{Keywords}
Innovation, diversity, knowledge spillovers, patents, startups, biotechnology.

\section{Introduction}
The contribution of industries to economic development varies significantly, and the emergence of the global biotechnology sector, which utilises living organisms and their compounds for diverse applications across industries, exemplifies this trend. The biotech sector in the US stands out as a remarkable success story, with revenues exceeding $10^5$ billion within just three decades \cite{biotech_US}. In the European context, the UK has gathered attention due to its position as the third-largest contributor to biomedical patents among 16 European countries. Additionally, the UK boasts the highest concentration of financially active biomedical startups and venture capital firms \cite{biotech_EU}.

Biotechnology has emerged as a critical driver of innovation in fields such as medicine, agriculture, and environmental sciences. However, the role of inventions and knowledge diversity in the success of biomedical startups remains unclear.
Understanding the dynamics and spatial patterns of biotech innovation is crucial for policymakers, entrepreneurs, and researchers aiming at fostering and supporting the growth of this sector. This study focuses specifically on the biotech landscape in the United Kingdom (UK) and examines the spaces of innovation and venture formation within the country.

The UK is recognised as a biotech hub, hosting numerous research institutions, universities, and industry players. It offers a unique ecosystem that fosters collaboration, knowledge exchange, and entrepreneurial activities. By analysing patent data, this study aims to gain insights into the spatial distribution of biotech innovation across UK, in order to identify the key regions and cities driving growth in this sector.

Innovation activity can be assessed through the analysis of patent data and technological advancements. Pugliese et al. demonstrated that technology serves as the most reliable predictor of industrial and scientific production in the coming decades \cite{pugliese2019unfolding}. The utilisation of patent data to monitor technological innovation is a well-established practice in academic research \cite{frietsch2010value, griliches1998patents, leydesdorff2015patents}. Thanks to the availability of different databases about patent documents and increased computational capabilities, patents have become a valuable resource for studying technological change \cite{youn2015invention}. 
Various entities, including academia (e.g., Hall et al. \cite{hall2001nber}), institutions (e.g., PATSTAT, REGPAT), and corporations (e.g., Google Patents), have contributed to the development of extensive collections of patent-related documents. This abundance of data has allowed researchers to explore multiple aspects of patented inventions, including their role in explaining the technological change, their interconnections, and their association with inventors and applicants \cite{youn2015invention, strumsky2011measuring, strumsky2012using}.
One notable characteristic of patent documents, particularly relevant for economic analysis, is the presence of codes associated with the claims made in patent applications. These codes categorise the scope of commercial rights sought by inventors. To facilitate an examination by patent office officials, claims are classified into technological areas using classification systems such as the IPC classification \cite{fall2003automated} or the United States Patent Classification (USPC) \cite{falasco2002bases, falasco2002united}. These classification systems employ hierarchical six-digit codes, which provide increasingly detailed definitions of technological domains. By mapping claims to classification codes, localised analysis of patents and patent applications within specific technology domains becomes possible.
However, it is essential to recognise the limitations of using patents as a proxy for measuring innovation \cite{hall2014choice}. Estimating the value of patents presents a significant challenge \cite{hall2005market}. While certain patents hold substantial market value, others may possess limited or no value. Furthermore, employing patent statistics as a comprehensive measure of economic and inventive activity is not without drawbacks \cite{pavitt1985patent, griliches1998patents}. It is crucial to acknowledge that inventions do not encompass all forms of knowledge production in the economy, and patents do not cover the entirety of knowledge generated \cite{arts2013inventions}. Additionally, patents represent just one among several indicators of knowledge and do not uniformly capture all sectors of the economy \cite{kogler2015intellectual, lanjouw1996innovation}. 

This study builds upon previous research that explored knowledge spillovers in the UK based on patent citations, with biotechnology showing a weaker effect compared to other technologies \cite{wilkinson2023KS}. By focusing on the local level, specifically the NUTS3 regions, and incorporating information on startups, we aim to address this limitation and investigate the influence of biotechnology incubators. Furthermore, we examine the regions in the UK that demonstrate high intellectual property (IP) potential and explore their capacity to drive knowledge accumulation in other industries.


\section{Data}

\subsection{Patents}

\begin{figure}
\centering
\subfloat[]
   {\includegraphics[width=.31\textwidth]{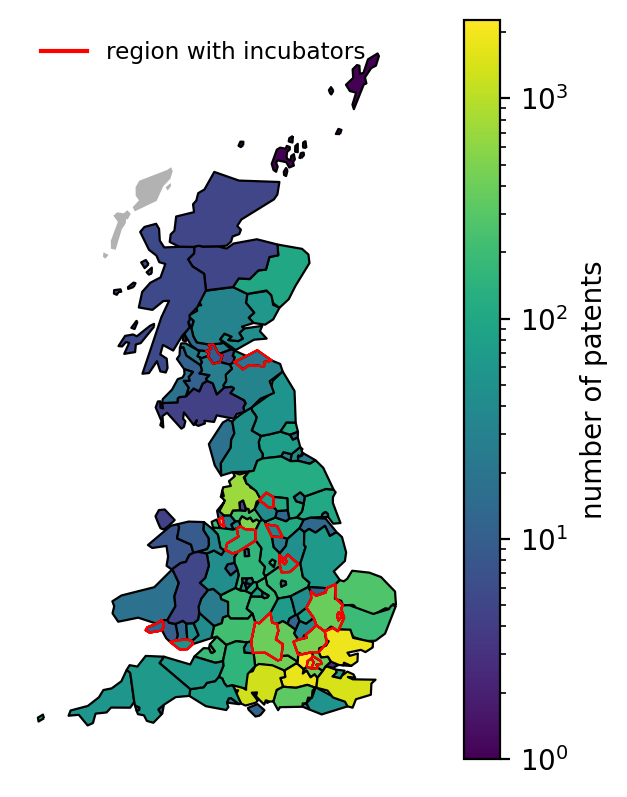}} \quad
\subfloat[]
   {\includegraphics[width=.31\textwidth]{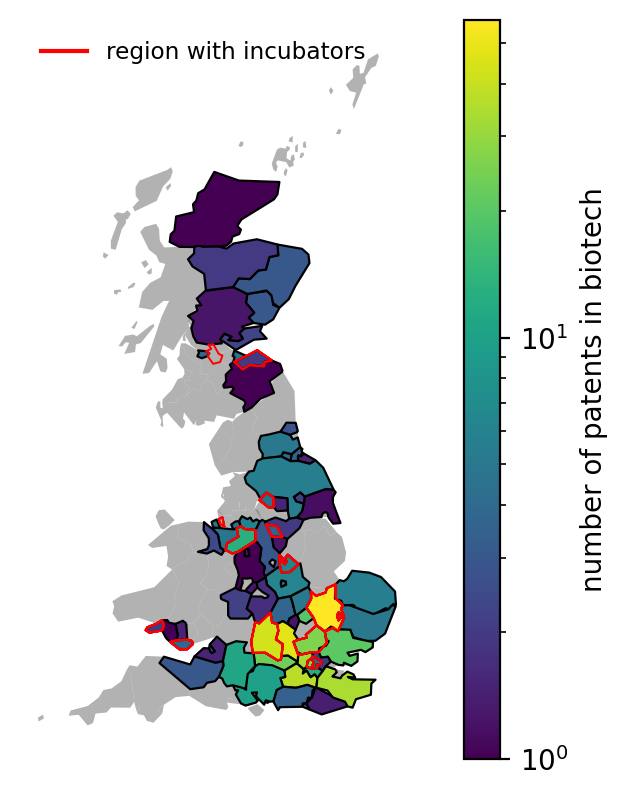}} \quad
\subfloat[]
   {\includegraphics[width=.32\textwidth]{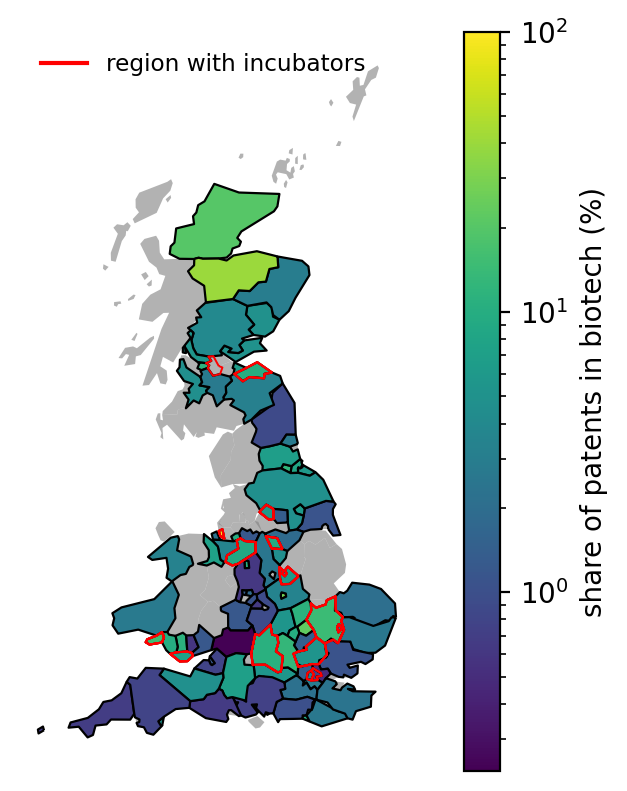}} 
\caption{\textbf{Geographical distribution of patent in the UK.} We plot the NUTS3 regions with red boundaries are those with incubators. \textbf{a}: Number of patents active in the UK. \textbf{b}: Number of these patents that belong to the biotech sector. \textbf{c}: Share, i.e. the percentage of patents which are in the biotech sector.}
\label{fig:patent}
\end{figure}

\paragraph{Sources:}
The Patent data used in this work is the same data as the one used in \cite{wilkinson2023KS}. It belongs to the OECD REGPAT database \cite{oecd_stat}, and it is from 1977 to 2019. It has been filtered such that only patents belonging to the UK, cited and citing, are considered. For further details on data manipulation please refer to \cite{wilkinson2023KS}. In that work, 43,751 total patents were considered in the study. We further filtered the data to consider only patents that have been cited at least once and that cite at least once, resulting in a total of 25,852 patents, from which 12,543 are cited at least once, and 15,745 cite at least once. 

\paragraph{Biotechnology patents:} Each patent in our dataset can be described by one or more technology codes (IPC codes), which provide information about the technology industry to which they belong. Patents that have at least one IPC belonging to the biotech classification are considered biotech. Selecting the biotechnology classification for the IPC codes, see Appendix, we identified 1,436 patents in this sector, from which 627 are cited at least once, and 937 cite at least once, see Fig.~\ref{fig:patent}.

\paragraph{Citations:}
The citation network, see Section \ref{sec:CN}, is derived from the citation dataset included in the OECD REGPAT database \cite{oecd_stat}. For this work, we excluded the patents that were outside of the UK citing other patents in the UK.

\paragraph{Geographical discrepancies:}
The UK patent database comprises patents from 1977 to 2018, encompassing a broad timeframe. Consequently, various patents are linked to different editions of NUTS3 available on the Eurostat website. To address this issue, all iterations of NUTS3 were downloaded, and the patents within each region were tallied. This approach ensures minimal overlap as the different NUTS3 versions primarily entail minor adjustments to the boundaries.

\subsection{Startups and Incubators}

\paragraph{Startups:}

\begin{figure}
\centering
\subfloat[]
   {\includegraphics[width=.30\textwidth]{figures/map_patents.png}} \quad
\subfloat[]
   {\includegraphics[width=.30\textwidth]{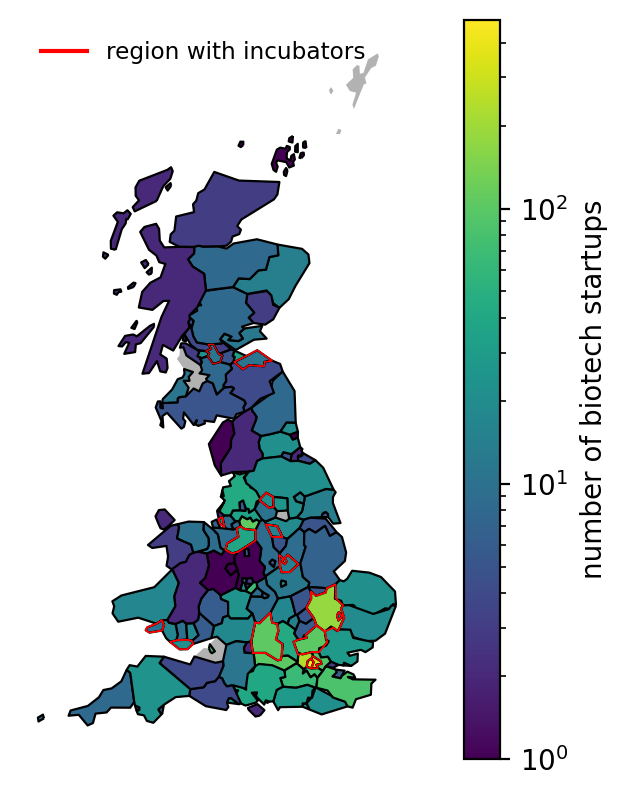}} \quad
\subfloat[]
   {\includegraphics[width=.33\textwidth]{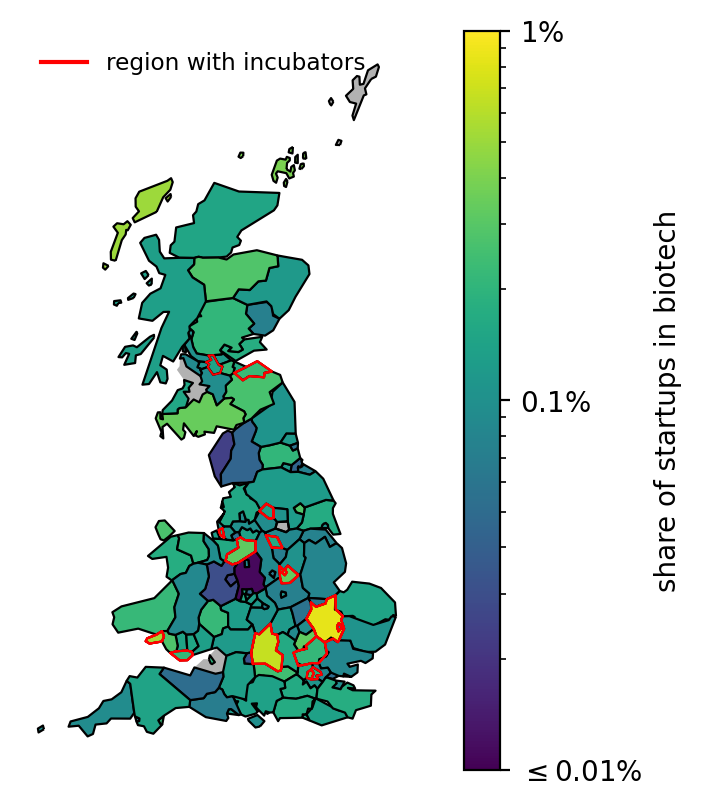}} 
\caption{\textbf{Geographical distribution of startups in the UK.} \textbf{a}: Number of startups active in 2018 in the UK. \textbf{b}: Number of these startups that are operating in the biotech sector. \textbf{c}: Share, i.e. the percentage of startups which are operating in the biotech sector. The NUTS3 regions with red boundaries are those with incubators.}
\label{fig:firm}
\end{figure}

While the term ``startup'' has become increasingly ubiquitous, a precise definition remains elusive due to its dynamic nature. A startup can be characterised as a young, innovative business. For the purpose of this study, we selected only those companies that have been registered for no longer than 5 years. This allowed us to define them as startups and choose them for further analysis. We extracted all new firms whose focus lies within the field of biotechnology. Biotechnology is a multidisciplinary field concerning many areas of society, including medicine, environmental science and many more, integrated with engineering sciences. In our search for startups, we referred to the official list \cite{sic_biotechnology} compiled by the government of the United Kingdom. This list consists of SIC codes which are used to classify businesses by industry in administrative units. The authors connected Biotechnology with Manufactures of Basic Pharmaceuticals, Pharmaceutical Preparations, Irradiation, Electromedical and Electrotherapeutic Equipment and Dental Equipment. The full list is available under the link \cite{sic_biotechnology}.

The firms' data has been extracted from Companies House \cite{CompHouse} for 2018, and we considered startups all firms that were created by 2014. Out of the total registered firms in 2018, around 51\% can be considered startups, leading to a total of 2,181,018. Out of those the share in biotech is 0.163\%, leading to a total of 3548. See Appendix for distribution. 

\paragraph{Incubators:} While a startup is typically considered a newly established business venture with a scalable business model and high growth potential, incubators, on the other hand, are organisations or programs designed to support and nurture startups during their early stages by providing resources, mentorship, and infrastructure. Technology incubators are established in order to promote the commercialisation of knowledge derived from the university-industry partnership and accelerate business development by providing access to seed investment \cite{business_incubators1}. The information about 20 biotechnology incubators in the UK was collected from \cite{business_incubators2}, including the geographic location and the institution that provided the platform (University-based, hospital-based, large pharma-based or stand-alone). For 13 incubators, we also collected information about their size and the number of tenant firms \cite{business_incubators1}.

\section{Methods}

\subsection{Citation network}
\label{sec:CN}
\begin{figure}[!ht]
    \centering
    \includegraphics[width=0.99\textwidth]
    {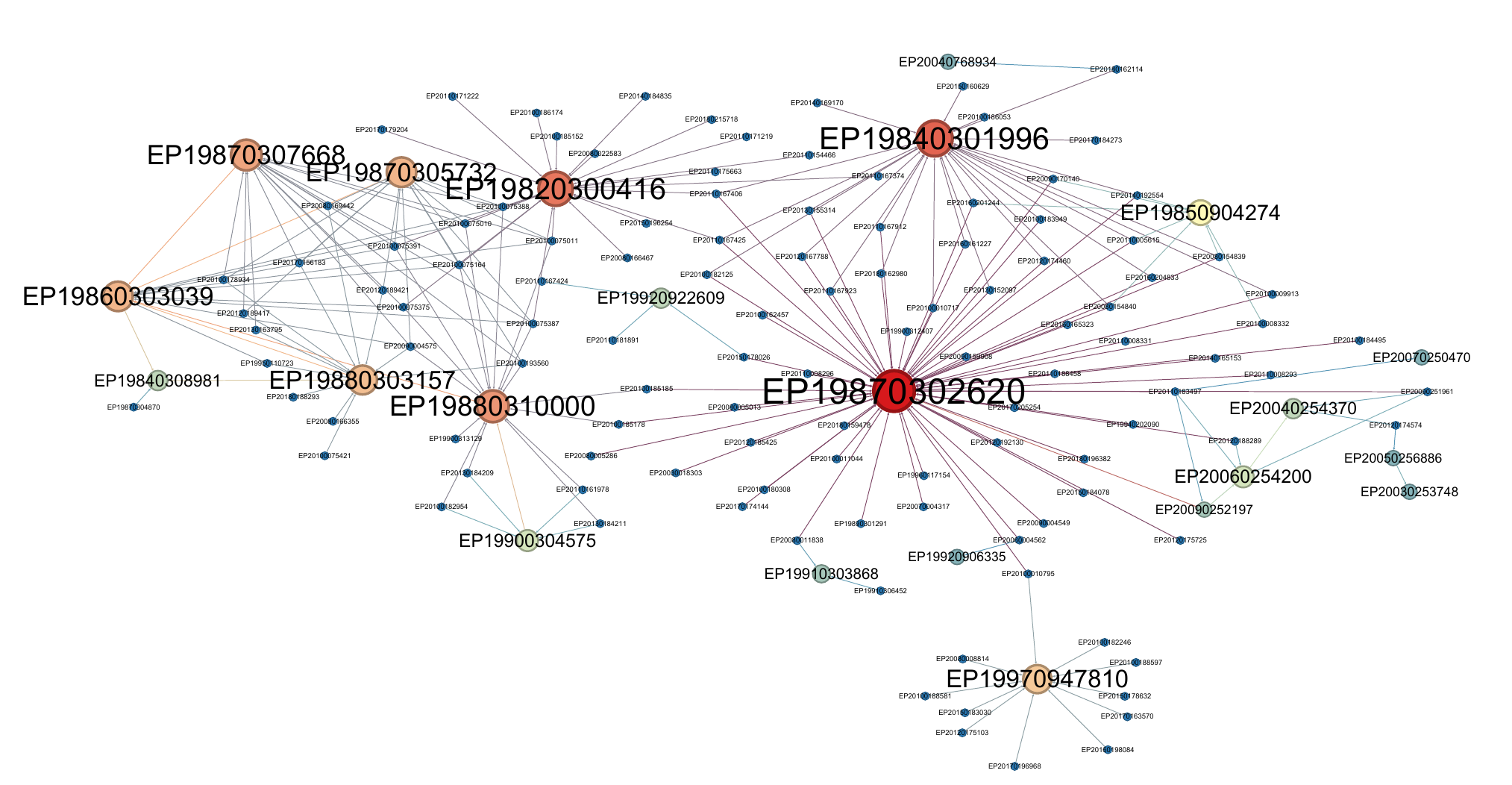}
    \caption{\textbf{Biotechnology patents}. An unweighted directed citation network of patents within the UK cited by other UK patents. The nodes' size and color (light to dark) are proportional to the in-degree of the node. For our study, the in-degree is a proxy of success.}
    \label{fig:citation_network}
\end{figure}

As mentioned previously, there are 15,745 patents that cite at least another patent. Using the unique identifiers for these patents we create a directed network. We show in Fig.\ref{fig:citation_network} the giant connected component (GCC) of this network after removing all nodes (patents) that are not from biotechnology. 



\subsection{Precursors of innovation and their diversity}

Diversity is considered an important driver for innovation. We will explore the diversity of the patents and that of the derived innovations from biotech, making use of a commonly employed measure of diversity \cite{cottineau2020nested}, Shannon's entropy. On the other hand, we will also explore whether there is an overlap between the different technologies involved in citing and cited patents using the Jaccard index.

\paragraph{Shannon's entropy:}
Shannon entropy (SE) \cite{shannon2001mathematical} measures the uncertainty or randomness in a dataset. It calculates the average amount of information or surprise in each data point. Higher entropy signifies more unpredictability, while lower entropy indicates more structure. SE is crucial in fields like data analysis, machine learning \cite{sharma2013classification} and cryptography \cite{cachin1997entropy} to assess dataset complexity and information content.
In this section, we utilise the SE to quantify the diversity of a patent. For each patent, the SE is defined as:
\begin{equation}\label{eq1}
SE = -\sum_{i}^{N} p_i \log(p_i)
\end{equation}
Here, $N$ represents the total number of unique IPC codes, and $p_i$ denotes the frequency of IPC technology $i$ within the patent, divided by the total number of unique codes in the dataset.

\paragraph{Technological similarity: } The Jaccard index \cite{jaccard1901distribution, costa2021further}, is a measure of the similarity between two sets. It is defined as the ratio of the size of the intersection of the sets to the size of their union. The Jaccard index ranges from 0 to 1, with 0 indicating no similarity and 1 indicating complete similarity between the sets. In our case, we compute the Jaccard index considering pair of patents $X$ and $Y$ with their set of IPC codes at 4-digit level $i$ and $j$. The calculation is computed with
\begin{equation}
    \text{Technological similarity}(X, Y) = \frac{| X_i \cap Y_j | }{| X_i \cup Y_j |}.
\end{equation}
Technological similarity $=1$ ($0$) for patents with identical (completely different) IPC codes.

\section{Results}

\subsection{Precursors to biotech innovation}

In order to foster innovation, it is important to understand which are the ideal conditions giving rise to the observed patents. In this section, we look at the precursors of innovation, which correspond to the cited patents and their IPC codes, and explore whether they belong or not to the same industry. We do this in a temporal manner, by looking at these over time.

\begin{figure}
\centering
\subfloat[]
   {\includegraphics[width=.47\textwidth]{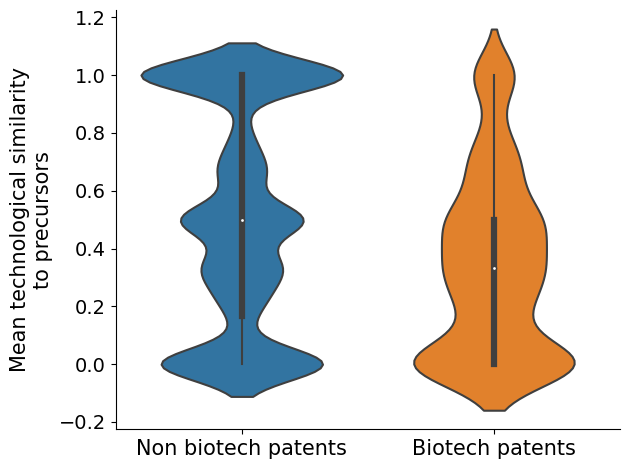}} \quad
\subfloat[]
   {\includegraphics[width=.47\textwidth]{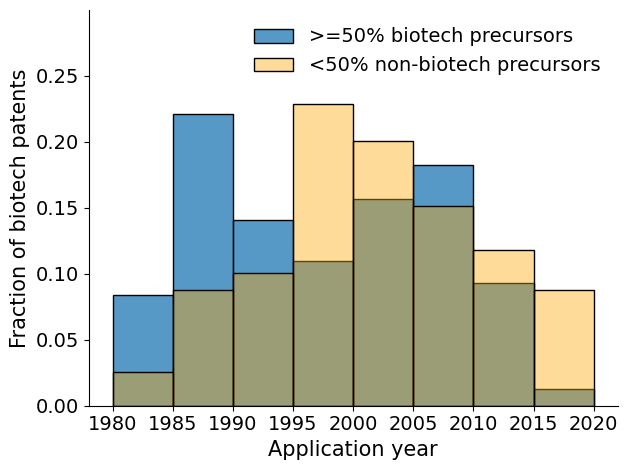}}
\caption{\textbf{Precursors of biotech innovations.} \textbf{(a)}: Mean technological similarity to precursors in biotech and non-biotech patents. \textbf{(b)}: Distribution of the biotech patents with and without biotech precursors over time.}
\label{fig:AM}
\end{figure}


In Fig. \ref{fig:AM} \textbf{(a)}, we plot the mean technological similarity of non-biotech and biotech patents to their precursors. From the large difference in the fraction of patents that are highly similar to their precursors, we see that innovations in biotech are more likely to come from different technologies than those outside of biotech.

As a further investigation, we check whether the precursors of biotech patents come from outside the biotech industry. Around 40\% of biotech patents have exclusively non-biotech precursors and around 55\% of them stem only from other biotech patents. In Fig. \ref{fig:AM}B, we plot the distribution of these two classes of biotech patents over time, finding that those with primarily non-biotech precursors are more recent than those with primarily biotech precursors.

\subsection{The role of incubators for innovation}

The UK has invested in biotech incubators across regions. The main role of these incubators, is to provide an ecosystem that supports biotech startups, by providing skills and expertise, to secure growth, and advance the industry. In this section, we explore whether the regions that contain these incubators, show an advantage with respect to regions that do not. We assess the impact of the incubators by looking at the startups and patents in biotech.

\begin{figure}[!ht]
    \centering
    \includegraphics[width=0.99\textwidth]{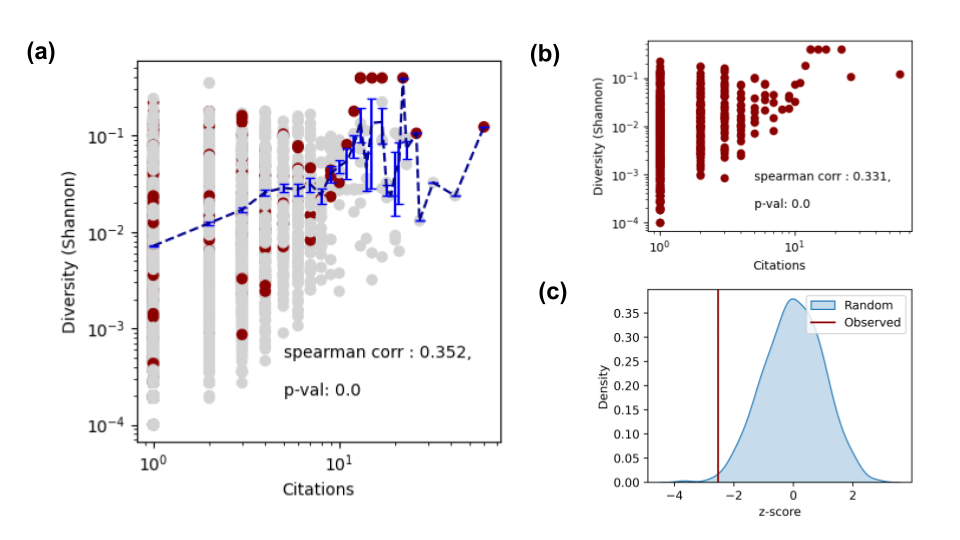}
    \caption{\textbf{Diversity vs Number of citations}. We plot the correlation between the diversity and the number of citing patents of all patents cited at least once. \textbf{(a)} All patents. \textbf{(b)} Biotech patents.
    The positive correlation indicates that highly cited patents are precursors to diverse innovations. \textbf{(c)} Comparing with the null model, the correlation is lower than expected.}
    \label{fig:entropy_msr}
\end{figure}

\begin{figure}[!ht]
    \centering
    \subfloat[]{\includegraphics[width=.475\textwidth]{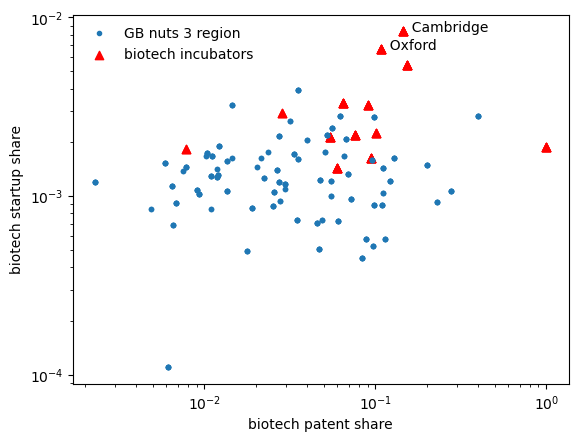}} \quad
    \subfloat[]{\includegraphics[width=.49\textwidth]{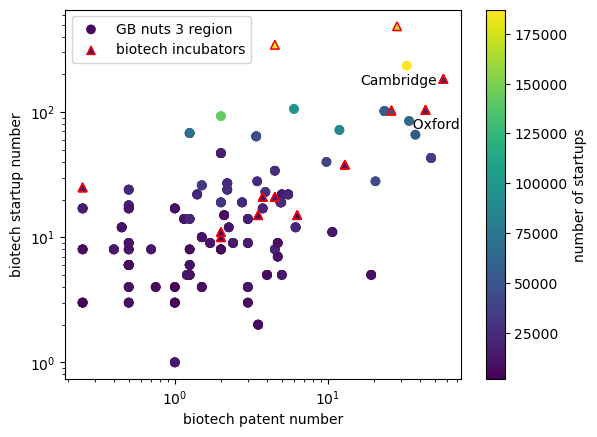}}
    
    \caption{\textbf{Geographical distribution of biotech activity.} Each dot represents a region (NUTS3 level), indicating its patent and startups share (a) and total (b) within biotechnology. Red triangles correspond to regions with biotech incubators, amongst which we highlight Oxford and Cambridge.} 
    \label{fig:scatter_share_patent_startups}
    
\end{figure}

\subsection{Derived innovations}

\subsubsection{Diversity}

\begin{figure}[!ht]
    \centering
{\includegraphics[width=.35\textwidth]{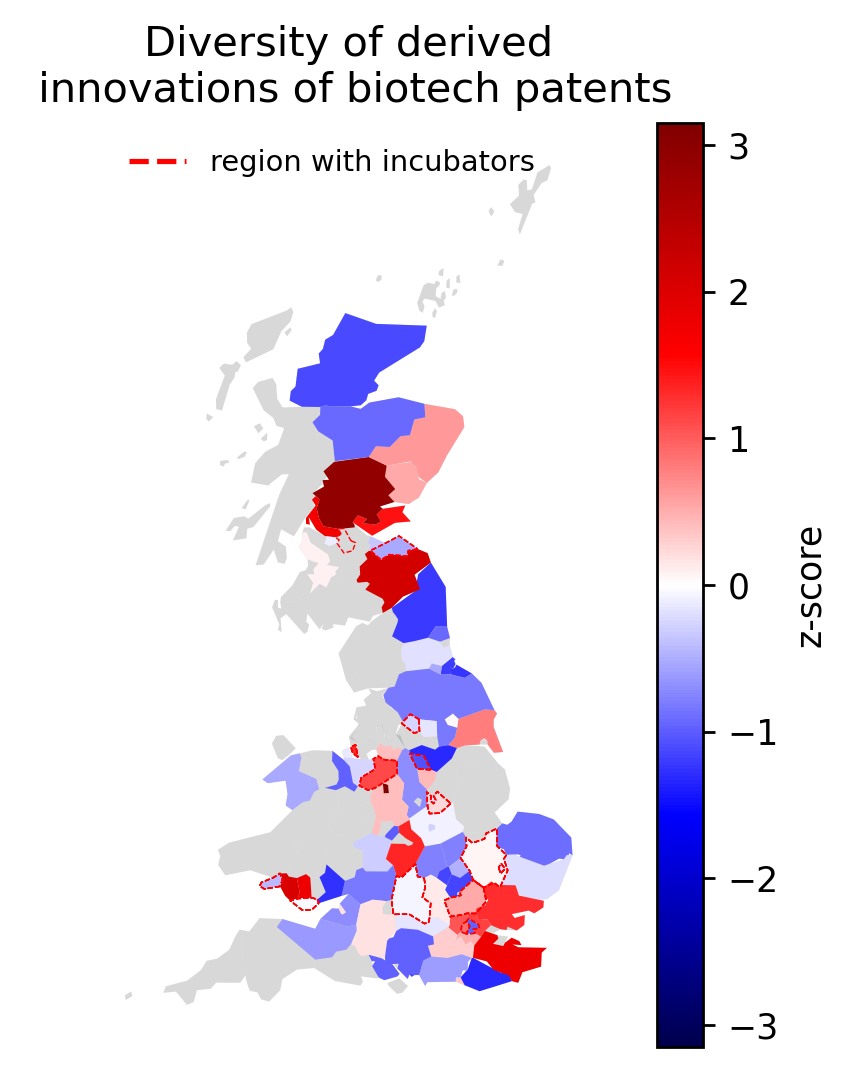}}

    \caption{\textbf{Average diversity of derived innovations of biotech patents} The z-score of the mean of the diversity (defined using Shannon Entropy) of innovations derived from the biotech patents in every NUTS3 region. The grey regions are those without any biotech patents. The regions with dashed boundaries are those containing incubators. Incubators do not own patents that crate more diverse innovations than other regions.} 
    \label{fig:share_patent_startups}

\end{figure}

Considering all patents cited at least once, we observe a positive Spearman correlation $= 0.352$ (p-value $<0.05$) between the number of citations received and the diversity of the citing patents (Fig.\ref{fig:entropy_msr}(a)). The corresponding correlation only for biotech patents is $0.331$ (p-value $<0.05$)(Fig.\ref{fig:entropy_msr}(b). Next, we define a simple null model for our citation network by using a directed configuration model \cite{hagberg2008exploring,newman2001random}. This preserves the degree sequence of the directed network, i.e. the total citations received by a patent. Comparing the observed value of the correlation with 1000 simulations of the null model, we observe that the correlation is less than expected (Fig.\ref{fig:entropy_msr}(c)). 

To check whether biotechnology patents lead to more diverse innovation, we classify all citing patents into those with (982) and without (14,763) biotech precursors. For each citing patent, we compute the mean technological similarity to its precursors and find that 30\% of the patents without biotech precursors are technologically identical to their precursors, while the same statistic for those with biotech precursors is 7.5\%. This indicates that biotech patents combine effectively with other patents to create novel innovations.

\section{Discussion}

Over the last 40 years, there was a huge number of biotechnological breakthroughs in the UK. The analysis of the patent citation network identified important innovations among which the most cited one is the fully humanised antibodies for therapeutic uses \cite{greg_winter}. We show that for biotech patents the mean technological similarity to innovation precursors is lower than average. However, in the last decade, the progress in the biotech industry is mostly driven by inventions from other fields (ecosystem-driven growth). The regional correlation between biotechnological patents and companies clearly highlights the importance of incubators, meant to facilitate knowledge exchange between science and industry. The most successful platforms are located in Oxford and Cambridge which were already well-established by the early 1990s \cite{clark_biotech_review}. Yet, the incubators themselves don't create much novelty for patents that use biotechnological advances on the regional level.

The technological diversity of patents that use biotechnological innovations strongly correlates with the importance of innovation. However, our null model suggests that the correlation is less than expected by random. This could be due to multiple reasons, firstly, by rewiring our network we lose the temporal structure of the network. Second, since the shuffling of edges does not take into account the geographic location of the patents when doing the randomisation, a patent can cite different patents anywhere in the UK. While this is possible, more often it is not what is observed. The influence of a patent is indeed driven by geographic proximity. To understand the effect of regional effects we look into the distribution of these patents and their effect geographical within the UK. 

Understanding the historical development of biotechnology as a newly emerged and rapidly evolving sector of the economy contribute towards the prioritisation of real economic goals. Under time and cost constraints, technology development analysis can positively affect policy-making and regulation. While the creation of biotechnological clusters positively affected economic growth in the past, the future biotech must promote the innovation diversity that will unlock equity and inclusive enterprise in the economy of the UK.

\section*{Acknowledgements}
This work is the output of the Complexity72h workshop, held at IFISC in Palma, Spain, 26-30 June 2023. \url{https://www.complexity72h.com/}. It means that after many coffees and laughs (and some beers) we came up with a plan for a future paper. This is the seed, the very beginning of a wonderful collaboration.


\section*{Appendix}
\subsection*{Biotechnological classification}\label{IPC_biotech}
List of IPC codes classified as biotech according to \cite{oecd_stat}: A01H1, A01H4, \\A01K67, A61K35/[12-79], A61K(38, 39), A61K48, C02F3/34, C07G(11, 13, 15), C07K(4, 14, 16, 17, 19), C12M, C12N, C12P, C12Q, C40B(10, 40/02-08, 50/06), \\ G01N27/327, G01N33/(53,54,55,57,68,74,76,78,88,92), G06F19/[10-18,20-24]

\subsection*{Startups}
\begin{figure}[!ht]
    \centering
    \includegraphics[width=1.0\textwidth]{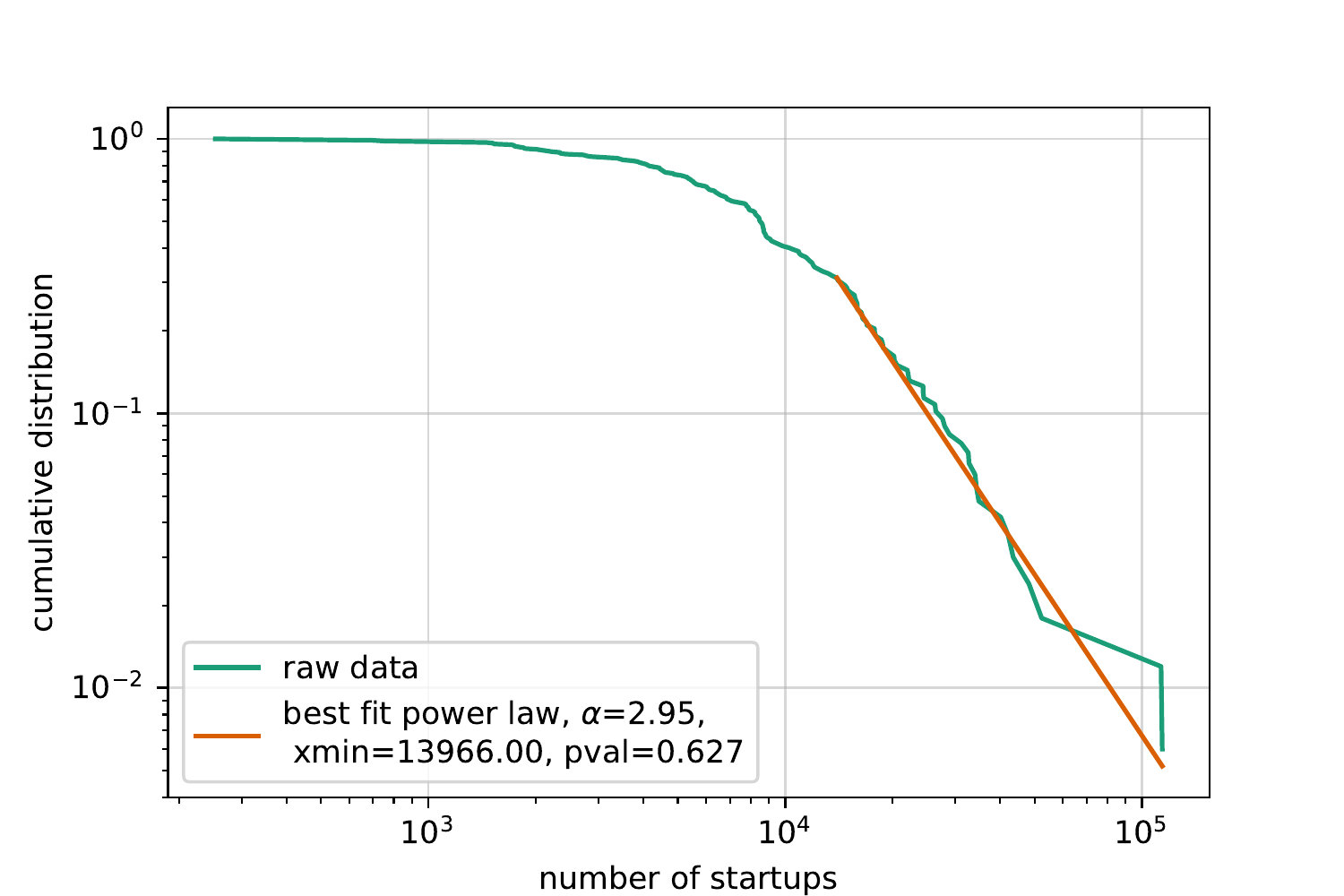}
    \caption{\textbf{Complementary cumulative distribution function of biotech startups}. The p-value from the Kolmogorov-Smirnov test suggests the tail of this distribution follows a power-law with an exponent $\alpha=2.95$.}
    \label{fig:distribution_startups}
\end{figure}

In the field of complex systems practice, it is common practice to fit various heavy-tailed distributions to real-world data. In figure \ref{fig:distribution_startups}, we can see the best-fit power law to the numbers of startups in the UK in different regions \footnote{For the purpose of the analysis we used the NUTS3 Territorial Units in the UK, and for each of them we computed the number of startups. See section \textbf{Startups and Incubators} for more details.}, using algorithms provided by Clauset et al. \cite{clauset_powerlaw}. As we can see, the number of startups seems to be following power law with an exponent of $\alpha=2.95$. Besides of visual confirmation, the Kolmogorov-Smirnov test \cite{KS_test} returned a p-value of $0.627$. It means that there is no reason to deny the hypothesis about the power law within this dataset.

However, it is not rare in the data analysis practice to consider only a truncated version of a given dataset. Here due to the recommendations in the publication of Clauset et al. \cite{clauset_powerlaw}, we focused solely on its right tail and performed a fitting procedure for values greater than 13 966. This means we kept about 30\%. However high p-value Kolmogorov-Smirnov suggest that there is indeed a power law. Direct conclusions of this are visible in the map in Figure \ref{fig:firm}(a), where one can see many green areas and only a few yellows. The occurrence of power law means that there are a few regions with an extraordinarily large number of startups.

\begin{figure}
    \centering
    \includegraphics[width=0.75\textwidth]{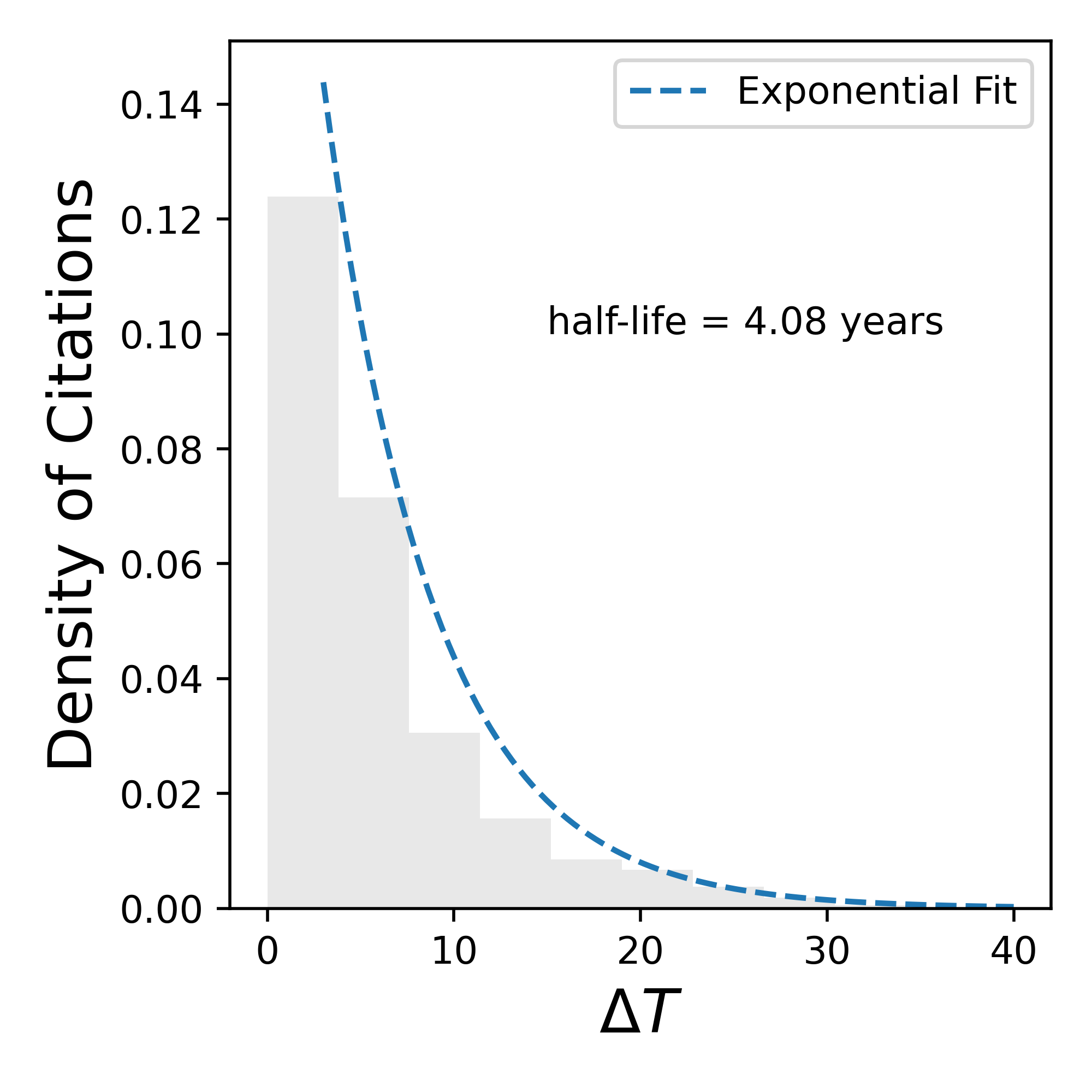}
    \caption{We show the histogram of the number of citations received by a patent after $\Delta T$ years of its appearance. The half-life of citations received for the patents in our study is $\approx$ 4 years. For the exponential fit, we use the function $N(t) = N_0 e^{-\lambda t}$ where N(t) are the citations received after $t$ years and fit for  $N_0$ and $\lambda$.}
    \label{fig:diversity_vs_Time}
\end{figure}

\end{document}


\title{\huge{Supplementary Information}\\\Large title}

\author{Io, me}

\date{}
\maketitle

\newpage

\tableofcontents
\newpage

\newpage

\section{Matching of postal code and NUTS3 UK regions.}

\newpage

\bibliography{bibliography}